\documentclass{ws-procs9x6}

\begin{document}

\title{Bo Andersson's contribution to experimental multiparticle production physics:\\an incomplete and biased selection\footnote{\uppercase{I}nvited talk to the commemoration of \uppercase{B}o \uppercase{A}ndersson in the \uppercase{XXXII I}nternational \uppercase{S}ymposium on \uppercase{M}ultiparticle \uppercase{D}ynamics; \uppercase{A}lushta (\uppercase{C}rimea), \uppercase{S}eptember 2002.}}

\author{ALESSANDRO DE ANGELIS}

\address{Dipartimento di Fisica dell'Universit\`a and INFN\\
Via delle Scienze 208, I-33100 Udine (Italy)\\
alessandro.de.angelis@cern.ch}


\maketitle

\abstracts{
Among the many topics related to soft QCD in which Bo Andersson gave an important contribution to experiments, this report selects a few for which the final word has not yet been pronounced.}

\section{Preamble}

It has not been easy to prepare a talk on Bo's contribution to experimental multiparticle production, since the memories related to work are closely mixed to personal experience. It is even more difficult to start writing these proceedings.

I knew Bo around 1990, immediately after the first runs of LEP. Bo came twice in Udine/Trieste for a visit, guested by INFN. We wrote just one paper together\cite{noibo}; it was about the difference in strenght of the Bose-Einstein Correlations (BEC) for particles belonging to gluon jets with respect to particles in quark jets. Today this subject would be ``\`a la page'', since the problem has been shown to be related to
the search for interconnection effects in W pairs. Our paper was not subitted to a journal, but published as an internal note of my experiment (DELPHI); in that paper, probably for the first time in the literature, a wine place (the {\em osmizza} of Monrupino, near Trieste) was acknowledged. 

I remember Bo for nice discussions, drinking together, for the original layout of his transparencies, for completely ignoring the time given to his talks, and for being able of disappearing from one place to reappear in a different one (like some saints).

Since summarizing the contribution of Bo to multiparticle 
dynamics would require a book (and actually he wrote one\cite{book}), I shall make a biased selection, being short on topics belonging to the past and concentrating more on topics from the future. During hist last years, in fact, Bo worked on topics for which the esperimental analysis is far from being completed.

\section{The Lund model and its impact on experimental HEP}
I shall start from a sentence by Feynman on QCD. ``We have a theory (...) so why can't we test it right away to see if it's right or wrong? Because what we have to do is calculate the consequences of the theory to test it. This time, the difficulty is this first step.''
Quarks and gluons, the elements of QCD, are not seen: only hadrons can be measured.
How to quantify hadronization? 
	The Lund string fragmentation model was the first answer to this question, and after three decades it is still the reference.

In the first fixed target experiments, the only way of measuring quantities was the direct one.
From the 80's, unfolding techniques became essential for extracting physical quantities from hadronic final states becoming more and more complicated due to the increase in energy:
a good modeling of the soft sector is the crucial point for the unfolding.

With the Lund Monte Carlo system\cite{lundmc}, based on the idea of Bo and of the Lund group,
the string-inspired MCs became soon a standard ``de facto'' for experimental physics
(several papers contain in the conclusion the 
sentence ``the experimental data are in agreement with the simulation'').
	
Many physical quantities have been measured by using Lund as a function of the unknown parameter, and then minimizing the $\chi^2$.
As a standard ``de facto'', Lund parameters were adjusted to the experimental data
(a huge computational program has been set up to tune the string parameters: I was supervising a thesis of computer science on this topic\cite{zanin}, and I heard of some others).

If you are always right, there is little to learn. Bo was thus always very attentive to
the indications of shortcomings coming from experiments,
especially in the extreme nonperturbative sector.
In the following I shall discuss a few topics where 
I am convinced that the suggestions from Bo can contribute to a better understanding of soft QCD, if followed.

\section{Miscellaneous topics where Bo's contribution is alive and kicking}

\subsection{BEC}
Models assume in general a hadron source
spherically symmetric in space.
Also in the most fundamental process, electron-positron annihilation, an ellipsoidal
structure fits more naturally the string model\cite{besphe}. The prediction of Bo was a ratio of approximately 0.5 between the short and the long axis; the data confirmed such an effect\cite{besphee}. Much has still to be done to implement a realistic simulation of
Bose-Einstein interference.

\subsection{Interconnection effects in W pairs}

When LEP2 started, the sector of ``soft QCD'' was activated mostly  by the problem of interconnection in WW events.

The problem can be stated as follows. W bosons are mostly produced in pairs at LEP2, and
each W has a probability about 2/3 of decaying into a quark and an antiquark. Since the lifetime of a W, from the Heisenberg principle, is one order of magnitude smaller than the hadronization time, one expects that when both Ws decay hadronically they cannot be treated as independent objects: their decay  diagrams are connected by (soft) gluons (Color Reconnection, CR)
and the hadrons in the final state are mixed together by ``exogamous'' BEC.

The problem is interesting {\em per se} and in connection with the determination of the W mass:
 the most accurate determination of this quantity can come from the hadronic WW channel, provided we understand interconnection.
 
Bo formulated a surprising (at least to me) prediction\cite{bopred}: interconnection in W pairs
could be be very small or zero, since each W hadronizes along a different string and thus they
are two separate objects.
This interpretation attributed a kind of reality of the string concept.
Since I am a partisan of the non-independence of the hadronization in the WW system, we had
long discussions on this topic; however, the final answer belongs to data.

As a matter of fact,  accourding to the data
Bo could be right for CR and for BEC as well\cite{intercon}: the LEP results are consistent with
the hypothesis of no interconnection at all. I cannot understand how Bo could be right; explaining this, however, would be rather his job than mine. 

\subsection{The baryon sector}

Bo was suggesting many tests of the Lund string  model in the baryon sector.
Fragmentation models contain several parameters which can be tuned according to data; a discrimination could be obtained by looking at 
				spin and angular momentum 
				correlations. 		
The know-how on the observables belongs mostly to the discussions between Bo and his colleagues.

\subsection{Production of prompt photons}
The decay of an unstable particle into charged particles can be thought as the sudden creation of rapidly moving charges. Such a variation of the electromagnetic field is accompanied by the emission of final state radiation. 
Experiments measure an order of magnitude more radiation than predicted\cite{rad}. Bo thought that this topic was very hot.

\section{A key role in the design of future experiments}

Last but not least, the
Lund Monte Carlo system 
is one of the keys for the next generation of experiments, from the project of the instruments to the definition of the trigger requirements and to the design of the software and of the analysis tools. Also in the most crucial analyses (hunting the Higgs), modeling the hadronization of very complex systems is the crucial point. 

To conclude, I think that without Bo's ideas soft QCD would be completely different today, and 
the design of future experiments would be much more problematic. In this sense I think that
most of Bo's contribution to experimental multiparticle dynamics is still to come.

\end{document}